\documentclass[aps,showpacs,twocolumn]{revtex4}
\usepackage{epsfig}
\usepackage{color}
\usepackage{amsmath}
\usepackage{amssymb}
\usepackage{bbm}

\newcommand{\be}{\begin{equation}}
\newcommand{\ee}{\end{equation}}
\newcommand{\bea}{\begin{eqnarray}}
\newcommand{\beaa}{\begin{eqnarray*}}
\newcommand{\eea}{\end{eqnarray}}
\newcommand{\eeaa}{\end{eqnarray*}}

\numberwithin{equation}{section}

\begin{document}

\title{\bf\Large {Curie temperature of the two band double exchange model for manganites}}

\author{Vasil Michev and Naoum Karchev\cite{byline}}

\affiliation{Department of Physics, University of Sofia, 1164 Sofia, Bulgaria}

\begin{abstract}
We consider two-band double exchange model and calculate the
critical temperature in ferromagnetic regime (Curie temperature).
The localized spins are represented in terms of the
Schwinger-bosons, and two spin-singlet Fermion operators are
introduced. In terms of the new Fermi fields the on-site Hund's
interactions are in a diagonal form and one accounts for them
exactly.  Integrating out the spin-singlet fermions we derive an
effective Heisenberg type model for a vector which describes the
local orientations of the total magnetization. The transversal
fluctuations of the vector are the true magnons in the theory, which
is a base for Curie temperature calculation. The critical
temperature is calculated employing the Schwinger-bosons mean-field
theory. While approximate, this technic of calculation captures the
essentials of the magnon fluctuations in the theory, and for $2D$
systems one obtains zero Curie temperature, in accordance with
Mermin-Wagner theorem.

\end{abstract}

\pacs{75.47.Lx, 71.27.+a, 75.30.Ds, 75.50.Cc}

\maketitle

\section {\bf Introduction}

The manganites are ones of the prominent representatives of strongly
correlated systems, where the effect of correlations among electrons
plays a crucial role. The growing interest on manganites is mainly
due to observation of resistivity changes by many orders of
magnitude upon the application of small magnetic fields, an effect
that carries the name of "Colossal Magnetoresistance".

The attempts to properly describe the physics of these materials by
means of simple perturbative methods typically fail \cite{Dagotto}.
One have to develop technic of calculation which capture the
essential features of the compounds. The double exchange model is a
widely used model for manganites \cite{cfm01,cfm20,Dagotto}. In
isolation, the ions of Mn  have an active $3d$-shell with five
degenerate levels. The degeneracy is presented due to rotational
invariance within angular momentum $l=2$ subspace. The crystal
environment results in a particular splitting of the five
$d$-orbitals  (\emph{crystal field spliting}) into two groups: the
$\emph{e}_g$ and $\emph{t}_{2g}$ states. The electrons from the
$\emph{e}_g$ sector, which form a doublet, are removed upon hole
doping. The $\emph{t}_{2g}$ electrons, which form a triplet, are not
affected by doping, and their population remains constant. The Hund
rule enforces alignment of the three $\emph{t}_{2g}$ spins into a
$S=3/2$ state. Then, the $\emph{t}_{2g}$ sector can be replaced by a
\emph{localized spin} at each manganese ion, reducing the complexity
of the original five orbital model. The only important interaction
between the two sectors is the Hund coupling between localized
$\emph{t}_{2g}$ spins and mobile $\emph{e}_g$ electrons. A strong
impact on the physics of manganites has the static Jahn-Teller
distortion which leads to a splitting of the degenerate $\emph{e}_g$
levels.

The double exchange model has a rich phase diagram, exhibiting a
variety of phases, with unusual ordering in the ground states. The
procedures followed to obtain the phase diagram in one band model
are different: numerical studies \cite{cfm31}, dynamical mean field
theory \cite{cfm32}, and analytical calculations \cite{cfm33,cfm34},
but four phases have been systematically observed: (i)
antiferromagnetism (AF) at a density of mobile electrons $n=1$, (ii)
ferromagnetism (FM) at intermediate electronic densities, (iii)
phase separation (PS) between FM and AF phases, and (iv) spin
incommensurable (IC) phase at large enough Hund coupling. The
competition between spin spiral incommensurate order or phase
separation and canted ferromagnetism is also a topic of intensive
study \cite{cfm33,cfm34,cfm35}. The phase diagram becomes more rich
if the orbital degeneracy is accounted for \cite{cfm36, cfm37}.

Double exchange model is also successful in calculating the critical
temperature in the ferromagnetic regime. Predictions about Curie
temperature in one band double exchange model are made using Monte
Carlo technique \cite{cfm31}, High-temperature series expansion
\cite{curie2}, Dynamical Mean-Field Approximation \cite{curie3} and
standard Mean-Field approach \cite{curie4}. The most striking
feature of the critical temperature as function of fermion density
is the symmetry with respect to $n=0.5$. The Curie temperature is
maximized at that point, and maximal value is different within
different approaches.

Very recently the critical temperature was calculated using two-band
model with account for the Jahn-Teller effect \cite{curie5,curie6}.
Again the characteristic feature of the temperature curves as a
function of charge carrier density is the symmetry with the respect
to $n=1$. The new assertion is that the Curie temperature increases
with increasing the interband hopping \cite{curie5}.

It is impossible to require the theoretically calculated Curie
temperature to be in accordance with experimental results. The
models are idealized, and they do not consider many important
effects: phonon modes, several types of disorder, Coulomb
interaction, etc. Because of that it is important to formulate
theoretical criteria for adequacy of the method of calculation. In
our opinion the calculations should be in accordance with
Mermin-Wagner theorem \cite{M-W}. It claims that in two dimension
there is not spontaneous magnetization at non-zero temperature.
Hence, the critical temperature should be equal to zero. It is well
known that Monte Carlo method of calculation does not satisfy this
criteria \cite{Dagotto}. It is difficult within Dynamical Mean-Field
theory to make a difference between two dimensional and three
dimensional systems. DEM is a good approximation when the
dimensionality goes to infinity. This made the analytical methods
important even for the assessment of the numerical results.

The paper is organized as follows: In Sec. II, starting from two
band double exchange model, we derive an effective Heisenberg like
model in terms of vector describing the local orientations of the
total magnetization. The transversal fluctuations of the vector are
the true magnons in the theory. This is a base for Curie temperature
calculation. Sec. III is devoted to phase diagrams of the model in
space of Hund's constant and charge carrier density. We calculate
spin-stiffness constant as a function of density which is important
step towards understanding the Curie temperature behavior as a
function of charge carrier density. The results for the Curie
temperature are reported in Sec. IV. A summary in Sec. V concludes
the paper. Spin-stiffness constant calculations are presented in the
Appendix.

\section {\bf Effective model}

We consider a two-band double-exchange model, with Hamiltonian: \bea
\label{H} &\displaystyle \nonumber H =- \sum\limits_{ll' < ij >
\sigma} \left( {t_{ll'}
c_{il'\sigma }^ +  c_{jl\sigma }  + h.c.} \right) -\\
&\displaystyle -2\sum\limits_{il} {J_l \vec S_i \cdot \vec s_{il} }
- \mu\sum\limits_{il} { c_{il\sigma }^ +  c_{il\sigma } }  \eea
where $l$,$l'$ are band indexes, $i$,$j$ are site labels, $\sigma$
are the spin indices, $c_{il\sigma }^ +$ and $c_{il\sigma }$ are
creation and destruction operators for mobile electrons, $\mu$ is
the chemical potential. $s_{il}$ is the spin of the conduction
electrons, and ${\bf S}_i$ is the spin of the localized electrons.
The sums are over all sites of a three-dimensional cubic lattice,
and $\langle i, j \rangle$ denotes the sum over the nearest
neighbors. We denote $t_{11}\equiv t_1$, $t_{22} \equiv t_2$, and
$t_{12}\equiv t_{21}\equiv t'$.

In terms of Schwinger-bosons ($\varphi_{i\sigma},
\varphi_{i\sigma}^{\dagger}$) the core spin operators have the
following representation: \be \label{sch} \vec S_i  =
\frac{1}{2}\varphi _{i\sigma }^ +  \vec t_{\sigma \sigma '} \varphi
_{i\sigma '} \qquad \qquad \qquad \varphi _{i\sigma }^ + \varphi
_{i\sigma }  = 2s \ee with the Pauli matrices
$(\tau^x,\tau^y,\tau^z)$.

The partition function can be written as a path integral over the
complex functions of the Matsubara time $\varphi_{i\sigma}(\tau)$
\,\, $(\varphi_{i\sigma}^+(\tau))$ and Grassmann functions
$c_{il\sigma } (\tau)$\,\,$(c^+_{il\sigma}(\tau))$. \be {\cal
Z}(\beta)\,=\,\int\,d\mu\left(\varphi^+,\varphi,c^+,c\right) e^{-S}
\label{cfm4} \ee with an action given by the expression \bea
\nonumber S & = & \int\limits^{\beta}_0 d\tau\left[
\sum\limits_i\left(\varphi^+_{i\sigma} (\tau)
\dot\varphi_{i\sigma}(\tau)+c^+_{il\sigma}(\tau)\dot
c_{il\sigma}(\tau)\right) + \right. \\ & & \left.
h\left(\varphi^+,\varphi,c^+,c\right)\,\right] \label{act} \eea
where $\beta$ is the inverse temperature and the Hamiltonian is
obtained from equations (\ref{H}) and (\ref{sch}) replacing the
operators with the functions.

We introduce spin-singlet Fermi fields \bea &&\hskip -1cm\Psi^A_{il}(\tau)=\frac {1}{\sqrt
{2s}}\varphi^+_{i\sigma}(\tau)c_{il\sigma}(\tau)\label{cfm8}\\
&& \hskip -1cm \Psi^B_{il}(\tau)=\frac {1}{\sqrt
{2s}}\left[\varphi_{i1}(\tau)c_{il2}(\tau)\,-
\,\varphi_{i2}(\tau)c_{il1}(\tau)\right] \label{cfm9} \eea which are
$U(1)$ gauge variant with charge -1 and 1 respectively \be
\label{cfm10} \hskip -.1cm \Psi'^A_{jl}(\tau)=e^{-i\alpha_j(\tau)}\Psi^A_{jl}(\tau)\quad
\Psi'^B_{jl}(\tau)=e^{i\alpha_j(\tau)}\Psi^B_{jl}(\tau)\ee The
equations (\ref{cfm8}) and (\ref{cfm9}) can be regarded as a SU(2)
transformation: \be \Psi_{il\sigma}  =
g_{i\sigma\sigma'}^+c_{il\sigma'}\,\, \Rightarrow \,\, g_i^+   =
\frac{1}{{\sqrt {2s} }}\left( {\begin{array}{*{20}c}
   {\varphi_{i1}^ +  } & {\varphi_{i2}^ +  }  \\
   { - \varphi_{i2} } & {\varphi_{i1} }  \\
\end{array}} \right)\label{cfm11}\ee
with $\Psi^A_{il}=\Psi_{il1}$ and $\Psi^B_{il}=\Psi_{il2}$.

In terms of the new Fermi fields, electron creation and destruction operators have the form:
\be\begin{array}{l}\begin{array}{l}
   {c_{il1}  = \frac{1}{{\sqrt {2s} }}\left( {\varphi _{i1} \Psi _{il}^A  - \varphi _{i2}^+ \Psi _{il}^B }
   \right)}\\[10pt]
   {c_{il2}  = \frac{1}{{\sqrt {2s} }}\left( {\varphi _{i2} \Psi _{il}^A  + \varphi _{i1}^+ \Psi _{il}^B } \right)}
\end{array}\\[20pt]
\begin{array}{l}
   {c_{il1}^ +   = \frac{1}{{\sqrt {2s} }}\left( {\varphi _{i1}^+ \Psi _{il}^{ + A}  - \varphi _{i2} \Psi _{il}^{ + B} }
   \right)}\\[10pt]
   {c_{il2}^ +   = \frac{1}{{\sqrt {2s} }}\left( {\varphi _{i2}^+ \Psi _{il}^{ + A}  +
   \varphi _{i1} \Psi _{il}^{ + B} } \right)}
\end{array}\end{array}\ee
and the spin of the conduction electrons ${\bf s}_{il}$ is
\be\label{cfm12} s_{il}^{\mu} = \frac 12 c^+_{il\sigma}
\tau^{\mu}_{\sigma\sigma'}c^{\phantom +}_{il\sigma'} = \frac 12{\rm
O_i}^{\mu \nu }\Psi_{il\sigma}^+
\tau^{\nu}_{\sigma\sigma'}\Psi_{il\sigma'}, \ee where
\be\label{cfm13} {\rm O_i}^{\mu \nu }  = \frac 12\mathop{Tr} g_i^ +
\tau ^\mu g_i\tau ^\nu.\ee It is convenient to introduce three basic
vectors which depend on the Schwinger-bosons \be\label{cfm14}
T^1_{i\mu}= {\rm O_i}^{\mu 1}\quad T^2_{i\mu}= {\rm O_i}^{\mu
2}\quad T^3_{i\mu}= {\rm O_i}^{\mu 3},\ee where ${\bf T}^3_{i}=
\frac 1s {\bf S}_i$. Then, the spin of the electrons can be
represented as a linear combination of three vectors ${\bf S}_{j}$,
${\bf P}_{j}={\bf T}^1_{j}+i{\bf T}^2_{j}$ and ${\bf P}^+_{j}={\bf
T}^1_{j}-i{\bf T}^2_{j}$ \bea\label{cfm15} {\bf s}_{il} & = &
\frac{1}{{2s}}{\bf S}_{i}\left( {\Psi^{A+}_{il}\Psi^A_{il}  - \Psi^{B+}_{il} \Psi^B_{il}} \right)\ \\
& + & \frac{1}{2} {\bf P}_{i} \Psi^{B+}_{il}\Psi^A_{il} +
\frac{1}{2}{\bf P}^+_{i} \Psi^{A+}_{il}\Psi^B_{il}.  \nonumber \eea
The basic vectors satisfy the relations ${\bf S}_i^2=s^2$, \,\,${\bf
P}^2_i={\bf P}^{+2}_i= {\bf S}_i\cdot{\bf P}_i= {\bf S}_i\cdot{\bf
P}^+_i=0$, and ${\bf P}^+_i\cdot{\bf P}_i=2$. Using the expression
(\ref{cfm15}) for the spin of itinerant electrons, the total spin of
the system  \be {\bf S}^{\rm tot}_{i} =  {\bf S}_i+{\bf s}_{i1}+{\bf
s}_{i2}\ee can be written in the form \bea {\bf S}^{\rm tot}_{i} & =
& \frac 1s  \left[s+\frac 12
\sum\limits_l\left (\Psi^{A+}_{il}\Psi^A_{il}-\Psi^{B+}_{il}\Psi^B_{il}\right)\right]{\bf S}_i\,+ \nonumber \\
& & \frac{1}{2} {\bf P}_i \sum\limits_l\Psi^{B+}_{il}\Psi^A_{il} + \frac{1}{2}{\bf
P}^+_i \sum\limits_l\Psi^{A+}_{il}\Psi^B_{il} \label{cfm16} \eea

The gauge invariance imposes the conditions \\
$\langle\Psi^{A+}_{il}\Psi^B_{il}\rangle=  \langle
\Psi^{B+}_{il}\Psi^A_{il} \rangle =0$. As a result, the
dimensionless magnetization per lattice site $\langle (S^{\rm
tot}_i)^z \rangle $ reads
\begin{equation}
\langle (S^{\rm tot}_i)^z \rangle =\frac 1s \left[s+\frac 12
\sum\limits_l \langle \left
(\Psi^{A+}_{il}\Psi^A_{il}-\Psi^{B+}_{il}\Psi^B_{il}\right) \rangle
\!\right] \langle {\bf S}^{z}_i \rangle  \label{cfm17}
\end{equation}

Let us average  the total spin of the system (Eq. \ref{cfm16}) in
the subspace of the itinerant electrons $  \langle {\bf S}^{\rm
tot}_i \rangle _f = {\bf M}_i$. The vector ${\bf M}_i$ identifies
the local orientation of the total magnetization. Accounting for the
gauge invariance, one obtains the following expression for ${\bf
M}_i$  \be\label{cfm19} \langle {\bf S}^{\rm tot}_i \rangle _f =
{\bf M}_i = \frac Ms {\bf S}_i \ee where
\begin{equation}
M = s+\frac 12 \sum\limits_l \langle \left
(\Psi^{A+}_{il}\Psi^A_{il}-\Psi^{B+}_{il}\Psi^B_{il}\right) \rangle
 \label{cfm17a}
\end{equation} can be thought of as an "effective spin" of the system $({\bf
M}_i^2=M^2)$.
 Now,
if we use Holstein-Primakoff representation for the vectors ${\bf
M}_j$ \bea\label{cfm20} & &
M_j^+ = M_{j1} + i M_{j2}=\sqrt {2M-a^+_ja_j}\,\,\,\,a_j \nonumber \\
& & M_j^- = M_{j1} - i M_{j2}=a^+_j\,\,\sqrt {2M-a^+_ja_j}
\\ & & M^3_j = M - a^+_ja_j \nonumber \eea the bose fields
$a_j$ and $a^+_j$ are the \textbf{true magnons} in the system. In
terms of the true magnons the Schwinger-bosons (\ref{sch}) have the
following representation \be\label{cfm21} \varphi_{i1} = \sqrt
{2s-\frac sM a^+_ia_i}, \qquad \varphi_{i2} = \sqrt {\frac sM}\,\,\,
a_i\,\,. \ee Replacing in Eqs. (\ref{cfm8}) and (\ref{cfm9}) for the
spin-singlet Fermions and keeping only the first two terms in $1/M$
expansion $\sqrt{1-\frac {1}{2M}\,\,a^+_ia_i}\simeq 1-\frac
{1}{4M}\,\,a^+_ia_i +\ldots $ we obtain \bea \Psi^A_{il} = c_{il1} +
\frac {1}{\sqrt{2M}}\,\,a^+_ic_{il2} - \frac {1}{4M}
a^+_ia_ic_{il1}+\ldots\label{cfm22}\\
\Psi^B_{il} =  c_{il2} - \frac {1}{\sqrt{2M}}\,\,a_ic_{il1} - \frac
{1}{4M} a^+_ia_ic_{il2}+\ldots\label{cfm23}\eea The equations
(\ref{cfm22}) and (\ref{cfm23}) show that the singlet fermions are
electrons dressed by a virtual cloud of repeatedly emitted and
reabsorbed magnons.

An important advantage of working with A and B fermions is the fact
that in terms of these spin-singlet fields the spin-fermion
interaction is in a diagonal form, the spin variables (magnons) are
removed, and one accounts for it exactly: \be \sum\limits_{il} \vec
S_i \cdot \vec s_{il}  = \frac{s}{2}\sum\limits_{il} \left( {\Psi
_{il}^{ + A} \Psi _{il}^A - \Psi _{il}^{ + B} \Psi _{il}^B } \right)
\ee

Replacing all this into the action (\ref{act}), we can rewrite it as
a function of the Schwinger-bosons and spin-singlet fermions. The
resulting action is quadratic with respect to the spin-singlet
fermions and one can integrate them out. The effective Hamiltonian
of the theory, in Gaussian approximation, is given by:
\be\label{eff} h_{\rm eff} = \rho \sum\limits_{  \langle  ij \rangle
}\left(a_i^+a_i + a_j^+a_j - a_i^+a_j - a_j^+a_i \right)\ee where
$\rho$ is the spin stiffness \eqref{rho}. Detailed calculation are
given in the appendix. Based on the rotational symmetry, one can
supplement the Hamiltonian (Eq. \ref{eff}) up to an effective
Heisenberg like Hamiltonian, written in terms of the vectors ${\bf
M_i}$ \be\label{cfm32} h_{\rm eff}= - J\sum\limits_{ \langle  ij
\rangle }{\bf M_i}\cdot{\bf M_j} \ee where $J=\rho /M$. The
ferromagnetic phase is stable if the effective exchange coupling
constant is positive $J > 0$.

It is important to highlight the difference between the above
effective theory \eqref{cfm32} and Ruderman-Kittel-Kasuya-Yosida
(RKKY) theory. The RKKY effective Hamiltonian is an effective
Heisenberg like Hamiltonian in terms of core spins ${\bf S}_i$,
obtained averaging in the subspace of the itinerant electrons
\cite{cfm41}. The subtle point is that if we use a
Holstein-Primakoff representation for the localized spins ${\bf
S}_i$, the creation and annihilation bose operators do not describe
the true magnon of the system \cite {cfm43}. The true magnons are
transversal fluctuations corresponding to the total magnetization
which includes both the spins of localized and delocalized
electrons. Therefore the RKKY validity condition requires small
Hund's coupling, and small density of charge carriers, which in turn
means that the magnetization of the mobile electrons is inessential.
In contrast of RKKY theory the effective model \eqref{cfm32} is
written in term of vectors $\textbf{M}_i$ which describe the local
orientations of the total magnetization, and the bose operators in
\eqref{cfm20} are the true magnons in the theory. This is essential
when one calculates the Curie temperature. The effective model
\eqref{cfm32} is obtained integrating out the spin-singlet fermions
\eqref{cfm8} and \eqref{cfm9}. In terms of these fermions the Hund's
interaction is in a diagonal form and we account for it exactly.
Hence, the effective theory \eqref{cfm32} is valid for arbitrary
values of Hund's constants and for all densities of charge carriers.

\section{\bf Phase diagrams}

Here we illustrate some of the features of our model, namely the
phase diagrams and how they change when we vary the model's
parameters. Since calculating $T_c$ is closely related to
calculating spin stiffness $\rho$, we have examined the behavior of
$\rho$ as a function of electron density in details.

\begin{figure}[!ht]
\epsfxsize=9cm 
\epsfbox{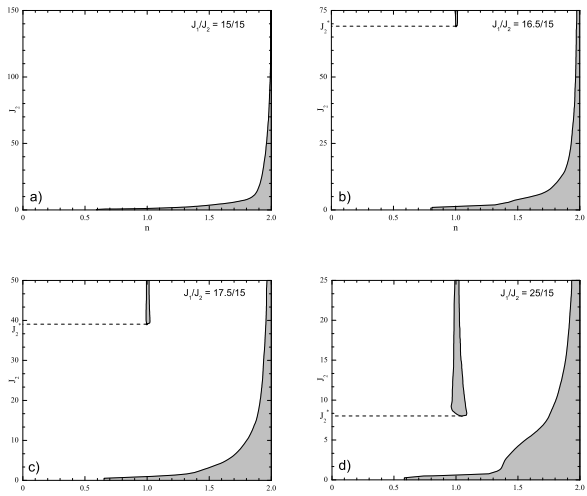} \caption{Phase diagrams for the four cases under consideration:
a) $J_1 = J_2=15$, b) $J_1=16.5,\, J_2 = 15$,
c) $J_1=17.5,\, J_2 = 15$ and d) $J_1=25,\, J_2 = 15$. White regions
correspond to FM phase, grayed ones to FM instability.}\label{fig1}
\end{figure}

The physics of the model depends on dimensionless parameters
$J_1/t_1$, $J_2/t_1$, $t_2/t_1$ and $t'/t_1$. Throughout this
chapter we fix the scale setting $t_1=t_2=1$, and use $J_1$, $J_2$
and $t'$ as a free parameters of the model. Also in this section
$t'=0.5$, and since we are describing manganite materials, we have
set $s=3/2$. The phase diagrams on Fig.\ref{fig1} are constructed by
plotting the curve $\rho = 0$ in coordinate system of carrier
density $n$ and $J_2/t_1$ for fixed ratio $J_1/J_2$. Regions where
$\rho >0$ correspond to the ferromagnetic phase (FM), while those
with $\rho <0$ are denoted here simply as non-FM ones, since
describing all possible phases is not the purpose of this paper.

We consider four different cases for the ratio $J_1/J_2$, namely
$J_1/J_2=15/15$, where the bands are degenerated, and three cases
with increasing bands' splitting $J_1/J_2 = 16.5/15$, $J_1/J_2 =
17.5/15$ and $J_1/J_2 = 25/15$.

When the bands are split, we observe an island of ferromagnetic
instability around the line $n=1$. The lowest point of the island is
denoted by $J^*_2$. Increasing the ratio $J_1/J_2$ increases the
island by lowering the value of $J^*_2$ (see Fig.\ref{fig1}). For
the degenerated bands ($J_1/J_2 =1$), we have not observed the
island of ferromagnetic instability, up to values as large as
$J_2=300$.

\begin{figure}[!h]
\epsfxsize=9cm 
\epsfbox{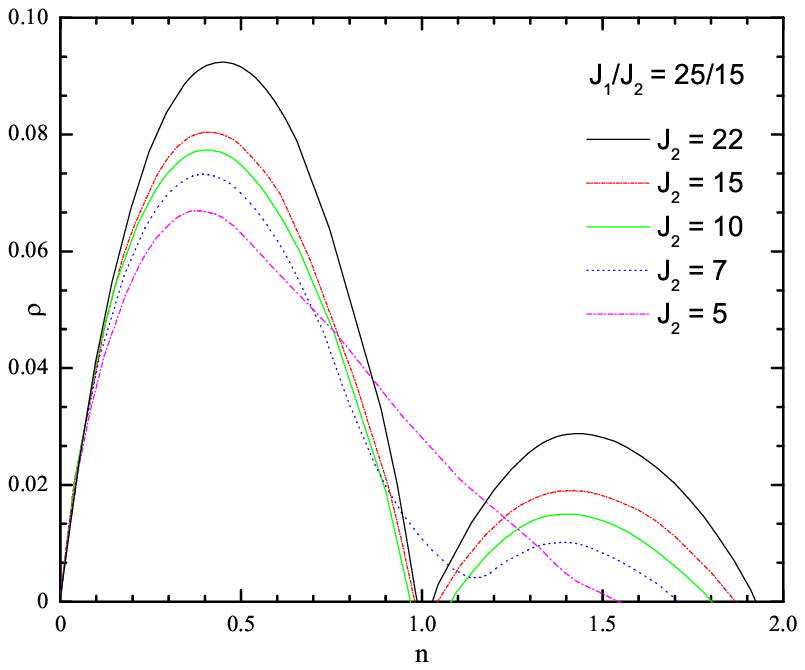} \caption{(color online) $\rho$ as a function of $n$ for the forth case $J_1/J_2=25/15$,
corresponding to fig. \ref{fig1}d.}\label{fig2}
\end{figure}

\begin{figure}[!h]
\epsfxsize=9cm 
\epsfbox{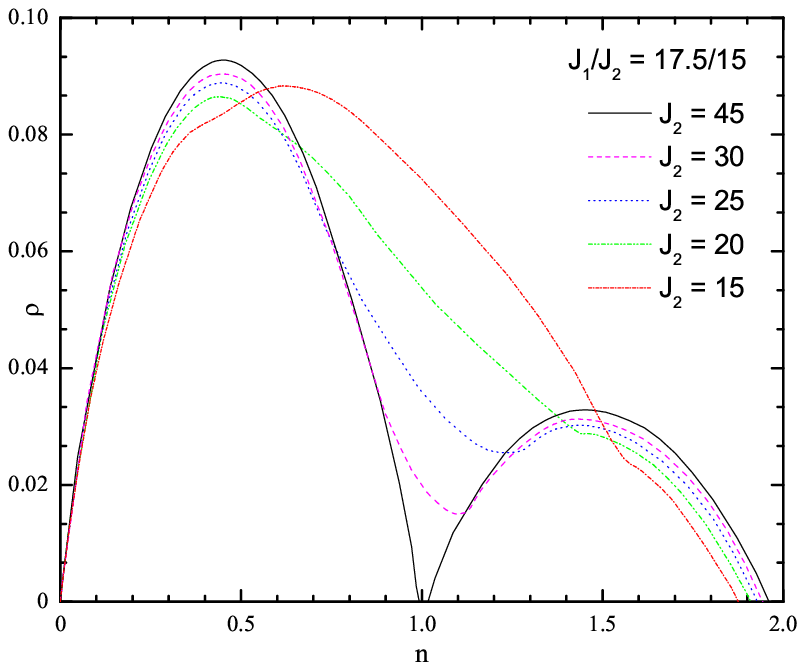} \caption{(color online) $\rho$ as a
function of $n$ for the third case $J_1/J_2=17.5/15$, corresponding to fig.
\ref{fig1}c.}\label{fig3} \end{figure}

The value of the spin stiffness constant $\rho$ depends on the point
($n,J_2$) in the phase diagram, for fixed ratio $J_1/J_2$. When the
point approaches the boundary of the ferromagnetic phase, the value
of $\rho$ decreases and reaches zero on the boundary (by the
definition of the boundary). We have calculated the spin stiffness
constant $\rho$ as a function of $n$, for fixed $J_2$ at zero
temperature. There are three distinctive cases: $J_2 > J^*_2$,
$J_2\lesssim J^*_2$ and $J_2 \ll J^*_2$.

\begin{figure}[!h]
\epsfxsize=9cm 
\epsfbox{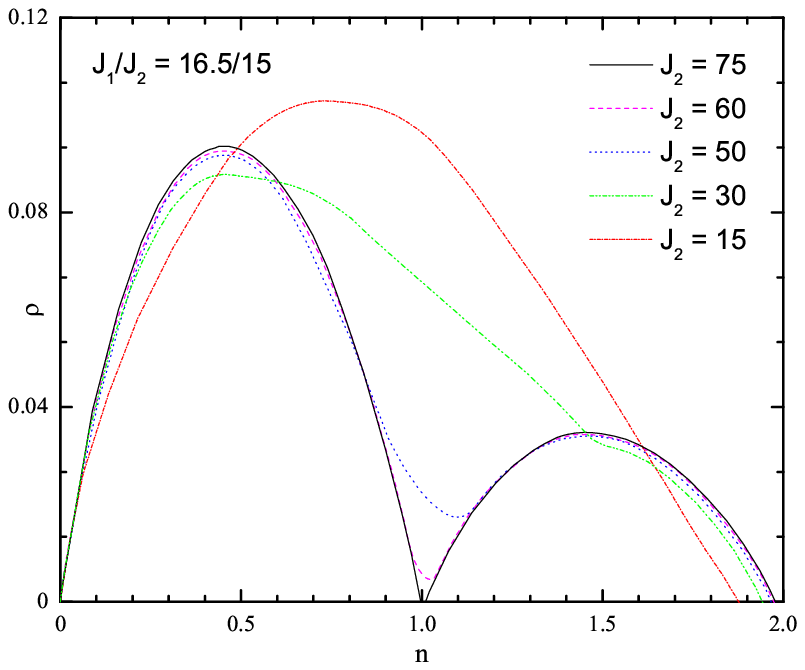} \caption{(color online) $\rho$ as a
function of $n$ for the second case $J_1/J_2=16.5/15$, corresponding
to fig. \ref{fig1}b.}\label{fig4}
\end{figure}

For the first one, the presence of ferromagnetic instability near
$n=1$ results in a function $\rho(n)$, which consists of two pieces,
one for $n$ in the interval $(0,1)$, and another for $n$ in the
interval $(1,2)$. For the ratio $J_1/J_2 = 25/15$ (Fig.\ref{fig2}),
such are the curves corresponding to $J_2=22$, $J_2=15$ and
$J_2=10$; for $J_1/J_2 = 17.5/15$ (Fig.\ref{fig3}) the curve
corresponding to $J_2=45$; and for $J_1/J_2 = 16.5/15$
(Fig.\ref{fig4}) the one corresponding to $J_2=75$. The important
characteristic of all these curves is the existence of two maxima,
one within interval $(0,1)$, and another one within interval
$(1,2)$. The global maximum is within the interval $(0,1)$, which is
result of the presence of two phase boundaries in the other
interval. In the case of degenerated bands, the absence of island of
instability leads to the absence of such type of function $\rho(n)$.

For the second one, $J_2$ is very close to $J^*_2$, hence near $n=1$
the spin stiffness constant is very small. As a result $\rho(n)$ is
a function with two maxima and one minimum. For the ratio $J_1/J_2 =
25/15$ (Fig.\ref{fig2}), such is the curve corresponding to $J_2=7$;
for $J_1/J_2 = 17.5/15$ (Fig.\ref{fig3}) the curves corresponding to
$J_2=30$ and $J_2=25$; and for $J_1/J_2 = 16.5/15$ (Fig.\ref{fig4})
the ones corresponding to $J_2=60$ and $J_2=50$. The minimal value
of the function decreases when $J_2$ approaches $J_2^*$. Again, in
the special case of degenerated bands, there is no such a curve.

\begin{figure}[!h]
\epsfxsize=9cm \epsfbox{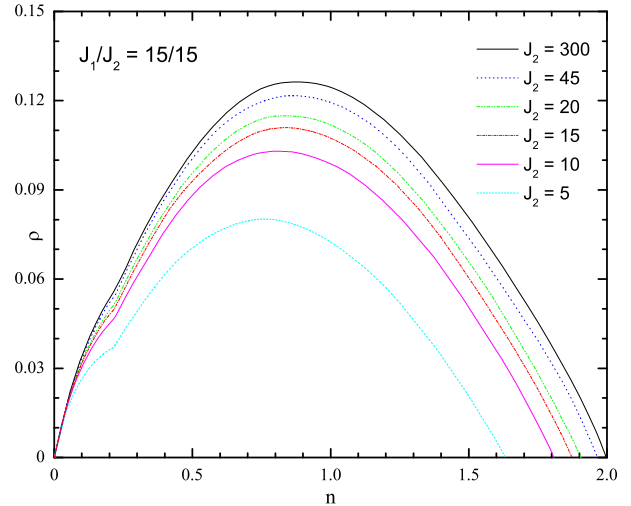} \caption{(color online) $\rho$ as a function of $n$
for the degenerated case $J_1/J_2=15/15$, corresponding to fig. \ref{fig1}a.}\label{fig5}
\end{figure}

For the third case, $J_2 \ll J^*_2$, we are well below the island of
ferromagnetic instability, and the function $\rho(n)$ has only one
maximum within the interval $(0,1)$. For the ratio $J_1/J_2 = 25/15$
(Fig.\ref{fig2}), such is the curve corresponding to $J_2=5$; for
$J_1/J_2 = 17.5/15$ (Fig.\ref{fig3}) the curves corresponding to
$J_2=20$ and $J_2=15$; and for $J_1/J_2 = 16.5/15$ (Fig.\ref{fig4})
the ones corresponding to $J_2=30$ and $J_2=15$. In the case of
degenerated bands, for all values of $J_2$ the curves are of this
type, because of the absence of instability island (Fig.\ref{fig5}).

It is widely known fact, that near $n=2$ the system is
ferromagnetically unstable and the spin stiffness constant
approaches zero when the carrier density approaches two. One can see
from the phase diagrams (Fig.\ref{fig1}), that with decreasing of
$J_2$, the ferromagnetic instability sets in for smaller value of
$n$. As a result we obtain that with decreasing $J_2$, the value of
$n$ for which $\rho(n)=0$ decreases. This is best observed in the
case of degenerated bands, where we have examined broader set of
values for $J_2$. With increasing of $J_2$, the point $\rho=0$ gets
closer to $n=2$ and the curve becomes more symmetric.

Since the maximum value of the Curie temperature corresponds to the
maximum value of the spin stiffness constant, it is important to see
how this value depends on the ratio $J_1/J_2$ for fixed $J_2$. We
choose the case $J_2=15$ (red lines in the figures), to compare our
results with the results in \cite{curie5}. Increasing $J_1/J_2$
increases the ferromagnetic instability, see phase portraits Fig.
\ref{fig1}, lowering the point $J_2^*$ which in turn leads to
decrease of the maximum value of $\rho$ and qualitative changes in
its behavior as a function of density $n$.

\section{\bf Curie temperatures}

To calculate the Curie temperature we utilize the Schwinger-bosons
mean-field theory \cite{S-b1,S-b2}. The advantage of this method of
calculation is that for $2D$ systems one obtains zero Curie
temperature, in accordance with Mermin-Wagner theorem \cite{M-W}.
So, while approximate, this technique of calculation captures the
essentials of the magnon fluctuations in the theory.

To proceed, we represent the effective spin vector $\textbf{M}_i$ by
means of Schwinger bosons $\phi_{i\sigma},\phi_{i\sigma}^+$, \be
M^{\nu}_i = \frac{1}{2}\sum\limits_{\sigma \sigma '} {\phi _{i\sigma
}^+ \tau^{\nu}_{\sigma \sigma '} } \phi _{i\sigma '} \qquad \phi
_{i\sigma}^+ \phi _{i\sigma} =2M \label{4.1} \ee

Next we use the identity \bea \nonumber \label{4.2} \textbf{M}_i
\cdot \textbf{M}_j  = \frac{1}{2}\left( {\phi _{i\sigma }^ +  \phi
_{j\sigma } } \right)\left( {\phi _{j\sigma '}^ +
\phi _{i\sigma '} } \right) -\\
-\frac{1}{4}\left( {\phi _{i\sigma }^ +  \phi _{i\sigma } }
\right)\left( {\phi _{j\sigma '}^ +  \phi _{j\sigma '} } \right)
\eea and rewrite the effective Hamiltonian in the form \be h_{\rm
eff}  =  - \frac{J}{2}\sum\limits_{ < ij > } {\left( {\phi _{i\sigma
}^ +  \phi _{j\sigma } } \right)\left( {\phi _{j\sigma '}^ +  \phi
_{i\sigma '} } \right)} \label{4.3} \ee where the second term in
\eqref{4.2} is equal to the constant $M^2$, because of the
constraint \eqref{4.1}, and we drop it. To ensure the constraint
\eqref{4.1} we introduce a parameter ($\lambda$) and add a new term
to the effective Hamiltonian \eqref{4.3} \be \label{4.4} \hat h
_{\rm eff} = h_{\rm eff}  + \lambda \sum\limits_i {\left( {\phi
_{i\sigma }^ + \phi _{i\sigma }  - 2M} \right)}\ee

We treat the four-boson interaction within Hartree-Fock
approximation. The Hartree-Fock hamiltonian which corresponds to the
effective hamiltonian \eqref{4.4} reads \bea \label{4.5} \nonumber
h_{\rm H - F}  = \frac{J}{2}\sum\limits_{ < ij > } {\bar u_{ij} }
u_{ij} &-& \frac{J}{2}\sum\limits_{ < ij > } {\left[ {\bar u_{ij}
\phi _{i\sigma }^ +
\phi _{j\sigma }  + u_{ij} \phi _{j\sigma }^ +  \phi _{i\sigma } } \right]}   \\
&+&\lambda \sum\limits_i {\left( {\phi _{i\sigma }^ +  \phi
_{i\sigma }  - 2M} \right)} \eea where $\bar u_{ij}\, ( u_{ij})$ are
Hartree-Fock parameters to be determined self-consistently. We are
interested in real parameters which do not depend on the lattice
sites, $u_{ij}=\bar u_{ij}=u$. Then in momentum space
representation, the Hamilonian \eqref{4.5} has the form \be
\label{4.6} h_{\rm H - F}  = \frac{{3J}}{2}Nu^2  - 2\lambda MN+
\sum\limits_k {{\varepsilon _k}\phi^+_k\phi_k } \ee where $N$ is the
number of lattice sites and \be \varepsilon _k = \lambda - Ju\left(
{\cos k_x + \cos k_y + \cos k_z } \right) \ee is the dispersion of
the $\phi_k$-boson (spinon).

The free energy of a theory with Hamiltonian \eqref{4.6} is \be F =
\frac{{3J}}{2}u^2  - 2\lambda M + \frac{{2T}}{N} \sum\limits_k {\ln
\left( {1 - e^{ - \frac{{\varepsilon _k }}{T}} } \right)} \ee where
$T$ is the temperature. The self-consistent equations for parameters
$u$ and $\lambda$ are \be \frac{{\partial F}}{{\partial u}} = 0
\qquad \frac{{\partial F}}{{\partial \lambda}} = 0 \ee We obtain a
system of two equations \bea \label{4.10}
&&\hskip -1cm u = \frac{2}{3}\frac{1}{N}\sum\limits_k {n_k \left( {\cos k_x  + \cos k_y  + \cos k_z } \right)}\\
&&\hskip -1cm M =\displaystyle \frac{1}{N}\sum\limits_k {n_k
}\label{4.11} \eea where $n_k$ is the bose function \be n_k =
\frac{1}{e^{\frac {\varepsilon_k} {T}}-1} \ee

To ensure correct definition of the bose theory \eqref{4.6} we have
to make some assumptions for the parameter $\lambda$. For that
purpose its convenient to represent it in the form \be \lambda  =
3Ju + \mu Ju \ee In terms of the new parameter $\mu$, the bose
dispersion is \be \varepsilon _k  = Ju\left( {3-\cos k_x  - \cos k_y
- \cos k_z+\mu} \right) \ee and the theory is well defined for $\mu
\geq 0$.

We find the parameters $\mu$ and $u$ solving the equations
(\ref{4.10}-\ref{4.11}). For high enough temperatures both
$\mu(T)>0$ and $u(T)>0$ and the excitation is gapped. It is the
spinon excitation in the theory in the paramagntic phase. Decreasing
the temperature leads to decrease of $\mu(T)$. At temperature $T_C$
it becomes equal to zero and long-range excitation emerges in the
spectrum. Hence the temperature at witch $\mu$ reaches zero is the
Curie temperature. We set $\mu=0$ in the system of equations
(\ref{4.10}-\ref{4.11}) and obtain a system of two equations for the
Curie temperature $T_C$ and the parameter $u$, which is the
renormalization of the exchange constant at Curie temperature. \bea
\nonumber &&\hskip -.5cm u = \frac{2}{3}\frac{1}{N}\sum\limits_k
\frac{\cos k_x  + \cos k_y  + \cos k_z }{\lower
5pt\hbox{$e^{\frac{\rho u}{M T_C }
\left(3 - \cos k_x  - \cos k_y  - \cos k_z  \right)}  - 1$}}  \\[5pt]
\\[-4pt]
&&\nonumber\hskip -.5cm M = \frac{1}{N}\sum\limits_k \frac{1}
{\lower 5pt\hbox{$e^{\frac{\rho u}{M T_C }\left( 3 - \cos k_x - \cos
k_y  - \cos k_z  \right)}  - 1$}}\eea

To calculate $T_C$ we solve the above system of equations, with
$M(T)$ and $\rho(T)$ calculated from equations \eqref{cfm17a} and
\eqref{rho} respectively.
\begin{figure}[!h]
\epsfxsize=9cm 
\epsfbox{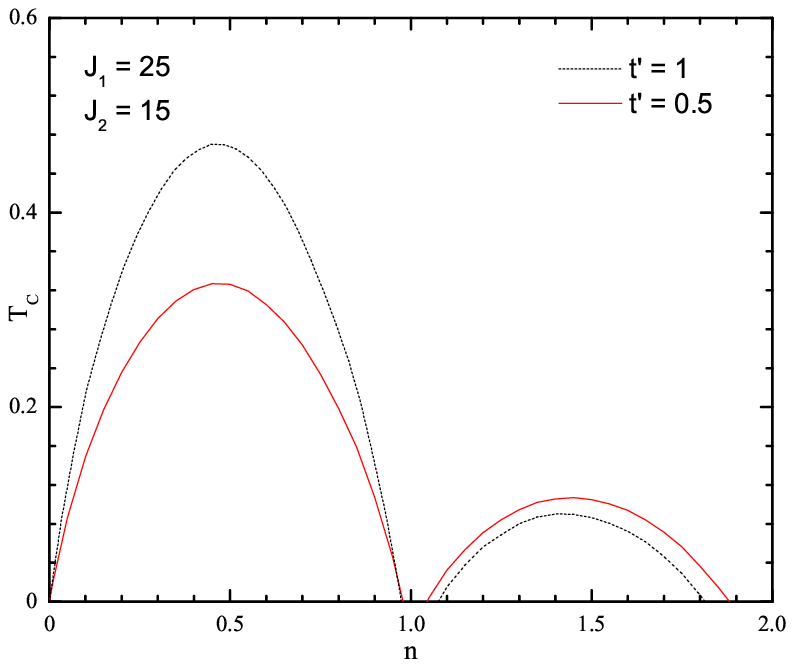} \caption{(color online) Curie temperature $T_C$
as a function of charge density $n$ for $J_1 = 25$, $J_2=15$, with
$t'=1/2$ and $t'=1$}\label{fig6}
\end{figure}

\begin{figure}[!h]
\epsfxsize=9cm 
\epsfbox{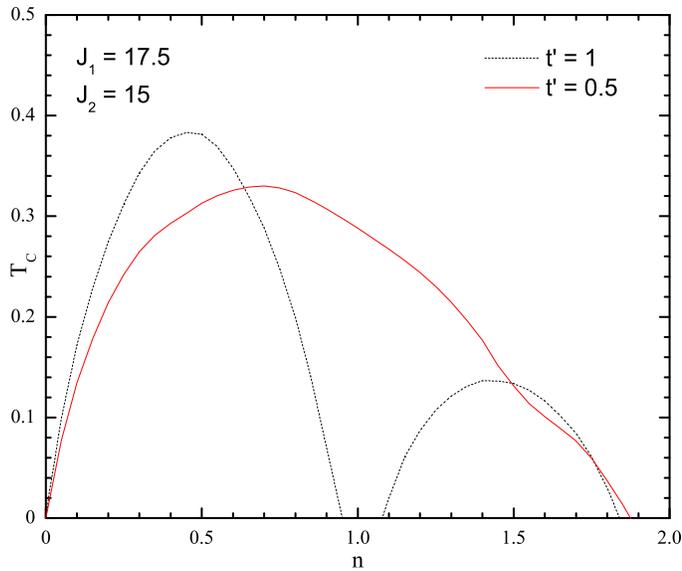} \caption{(color online) Curie temperature
$T_C$ as a function of charge density $n$ for $J_1 = 17.5$,
$J_2=15$, with $t'=1/2$ and $t'=1$}\label{fig7}
\end{figure}

\begin{figure}[!th]
\epsfxsize=9cm 
\epsfbox{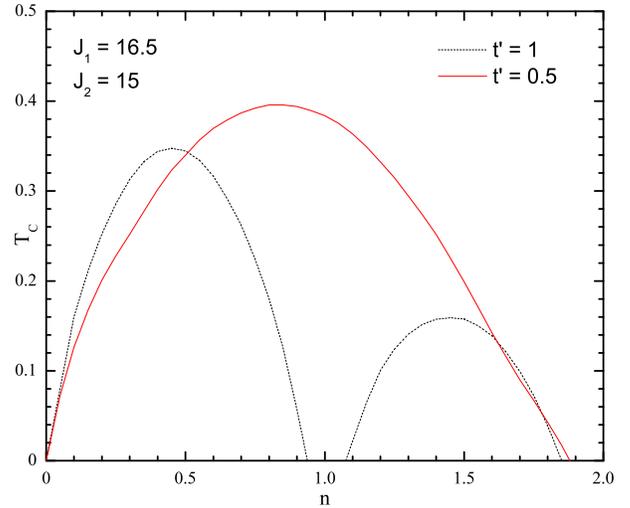} \caption{(color online) Curie temperature
$T_C$ as a function of charge density $n$ for $J_1 = 16.5$,
$J_2=15$, with $t'=1/2$ and $t'=1$}\label{fig8}
\end{figure}

\begin{figure}[!h]
\epsfxsize=9cm 
\epsfbox{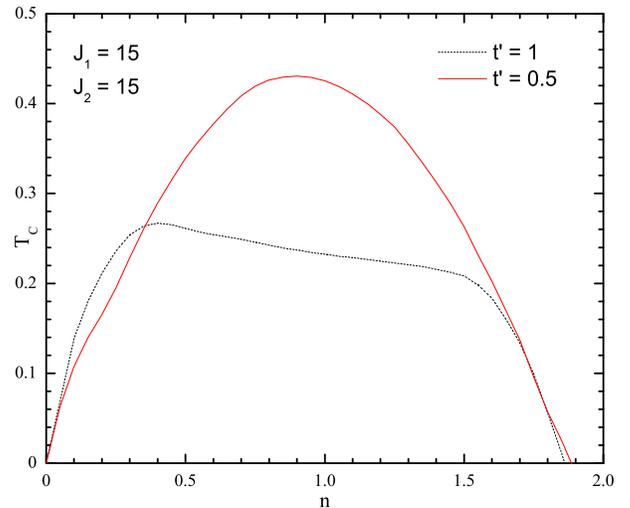} \caption{(color online) Curie temperature $T_C$
as a function of charge density $n$ for the degenerated case $J_1 =
15$, $J_2=15$, with $t'=1/2$ and $t'=1$}\label{fig9}
\end{figure}

The results for the Curie temperature $T_C$ as a function of charge
density $n$ are plotted on figures \ref{fig6}-\ref{fig9}, for two
different values of interband hopping parameter $t'$. In all cases
we have set $t_1=t_2=1$, and consider a theory with $s=3/2$ for the
core spins. We want to compare our results with the results in
\cite{curie5}, so we have set $J_2 = 15$ and $t'=0.5$. Since our
method of calculating $T_C$ involves $\rho(T)$, the resulting curves
are very similar to the ones in section III
(Fig.\ref{fig2}-\ref{fig5}), and bear their characteristics. As
above we have three different choices for the parameters: $J_2>
J^*_2$, $J_2\lesssim J^*_2$ and $J_2 \ll J^*_2$. For the biggest
ratio we consider $J_1/J_2 = 25/15$ (see Fig.\ref{fig6}), we have a
two-piece function, since our chosen value $J_2 > J_2^*$. With
decreasing the ratio, the point $J_2^*$ moves to higher values and
$J_2$ becomes smaller than $J^*_2$. As a result the curves we
obtained have only one maximum. It is important to note, that even
for the degenerated case $J_1/J_2=1$, the point at which $T_C$
reaches zero is smaller than $n=2$ unlike the results in
\cite{curie5}. This can be also seen from the phase portraits
(Fig.\ref{fig1}), where the instability of ferromagnetism near the
charge carriers density $n=2$ is evident. One can also note that
decreasing the ratio $J_1/J_2$ leads to increase in the maximum
value of $T_C(n)$.

The most important difference between our curves of critical
temperature as a function of $n$ and the ones in \cite{curie5} is
the lack of symmetry with respect to $n=1$. This asymmetry
originates from asymmetry in the phase diagrams, which in turn leads
to asymmetry of the spin-stiffness curves.

We have also examined the effect of $t'$. Increasing its value
results in enlargement of the ferromagnetic instability island
around $n=1$. The value of $J^{*}_2$ decreases and the width of the
island increases. This in turn leads to both quantitative and
qualitative changes in $T_C(n)$ (see the black lines of
Fig.\ref{fig6}-\ref{fig9}). An important conclusion is that Curie
temperature increases with increasing $t'$ if the bands are strongly
split (Fig.\ref{fig6}-\ref{fig7}), while for weakly split bands we
obtained an opposite behavior, the critical temperature decreases
(Fig.\ref{fig8}-\ref{fig9}).

\section{Summary}

In summary, we have calculated the Curie temperature in two-band
double exchange model. First we reduced the model to an effective
Heisenberg type model for a vector which describes the local
orientations of the total magnetization. Next, we use
Schwinger-bosons mean-field theory to calculate the critical
temperature. This technique of calculation is in agreement with
Mermin-Wagner theorem, which means that employing our method of
calculations for $2D$ system one obtains $T_C=0$ \cite{S-b1,S-b2}.

There are many quantitative and qualitative differences between our
results and the results obtained within Dynamical Mean-Field Theory
or Monte Carlo simulation approach. Maybe the most significant
difference is that DMF and MC calculations lead to temperature's
curves, as a function of fermion density, symmetric with respect to
$n=1$. This is not the result in the present paper. The asymmetry of
the $T_C(n)$ curves with respect to $n=1$ is a characteristic
feature in our approach. This asymmetry is seen looking at the phase
diagrams. It predetermines the asymmetry of the spin-stiffness
curves which lead directly to the asymmetric $T_C(n)$ curves. The
symmetry mentioned in the paper \cite{curie5} is possible if the two
bands are degenerated and Hund's constants are unphysically large.

Another important result reported in previous papers is that the
critical temperature increases with increasing the interband
hopping. Our calculations show that this is true when the band
splitting is strong. If the bands split weakly, the assertion is
opposite. The Curie temperature increases when the interband hopping
decreases.

\section{Acknowledgments}

The authors acknowledge the financial support of the Sofia
University, Project 037/2007

\appendix
\section{Calculation of \mbox{\Large $\rho$}}

Here we present a detailed derivation of the spin stiffness constant
$\rho$. We start from the two-band hamiltonian \eqref{H} and rewrite
it in terms of Scwinger bosons \eqref{sch} and spin-singlet fermions
(\ref{cfm8}-\ref{cfm9}).

The resulting action is quadratic with respect to the spin-singlet
fermions and one can integrate them out. To do so, it is convenient
to represent the action as a sum of three terms: \be S = S_{\rm f} +
S_{\rm s-f}^{(1)} +S_{\rm s-f}^{(2)}\ee where $S_{\rm f}$ is the
free fermion action: \be \hskip -.2 cm S_{\rm f} =
\!\int\limits_0^\beta \!d\tau \Bigg\{\!\! \sum\limits_{i\sigma}\!
\left(\! \Psi_{i1\sigma}^+ \frac{\partial}
{\partial\tau}\Psi_{i1\sigma} + \Psi_{i2\sigma}^+
\frac{\partial}{\partial\tau}\Psi_{i2\sigma}\! \right)\!+\! H_{\rm
f}\!\Bigg\}\ee with free fermion hamiltonian $H_{\rm f}$ \bea H_{\rm
f} = -s\sum\limits_{il} {J_l \left( {\Psi_{il}^{+A}\Psi_{il}^A -\Psi
_{il}^{ + B} \Psi _{il}^B } \right)}- \mu \sum\limits_{il\sigma}
{\Psi _{il\sigma}^+ \Psi _{il\sigma} }\nonumber\\[5pt]
-\nonumber \sum\limits_{ < ij >\sigma}\! {t_1 }\! \left( {\Psi
_{i1\sigma}^+  \Psi _{j 1\sigma}  + \Psi _{j1\sigma}^+  \Psi _{i1\sigma} } \right)\\[-4pt]
\\
- \sum\limits_{ < ij >\sigma }\! {t_2 }\! \left(
{\Psi _{i 2\sigma}^+ \Psi _{j2\sigma}  + \Psi _{j 2\sigma}^+\Psi _{i2\sigma} } \right)\nonumber\\
\hskip -.5cm -\!\!\sum\limits_{ < ij >\sigma}\!\!{t' }\! \left(
{\Psi _{i1\sigma}^+ \Psi _{j 2\sigma} + \Psi _{j2\sigma}^+  \Psi _{i
1\sigma} + \Psi _{i2\sigma}^+ \Psi _{j 1\sigma}  + \Psi
_{j1\sigma}^+ \Psi _{i2\sigma} } \right)\nonumber \eea We remind of
the notations $\Psi _{il1}=\Psi _{il}^A$ and $\Psi _{il2}=\Psi
_{il}^B$, so that the sum over $\sigma$ in the above equation is a
sum over $A$ and $B$. It is convenient to represent the term
describing spin-fermion interaction as a sum of two terms:
\begin{widetext}\bea \label{S1}\nonumber S_{\rm s-f}^{(1)} =\int\limits_{0}^\beta d\tau
\bigg\{ -\frac{t_1}{2s}\sum\limits_{< ij> }  \bigg[\left( {\varphi_{i\sigma}^+ \varphi_{j\sigma}-2s} \right)
\left( {\Psi _{1i}^{ + A} \Psi
_{1j}^A  + \Psi _{1j}^{ + B} \Psi _{1i}^B } \right) + \left(
{\varphi_{j\sigma}^+ \varphi_{i\sigma}-2s} \right)\left( {\Psi
_{1i}^{+ B} \Psi _{1j}^B  + \Psi _{1j}^{ + A} \Psi _{1i}^A } \right)\bigg]\\
-\frac {t_2}{2s}\sum\limits_{< ij> }  \bigg[\left(
{\varphi_{i\sigma}^+ \varphi_{j\sigma}-2s} \right)\left( {\Psi
_{2i}^{ + A} \Psi _{2j}^A  + \Psi _{2j}^{ + B} \Psi _{2i}^B }
\right) + \left( {\varphi_{j\sigma}^+ \varphi_{i\sigma}-2s}
\right)\left( {\Psi _{2i}^{+ B} \Psi _{2j}^B  + \Psi _{2j}^{ + A} \Psi _{2i}^A } \right)\bigg] \nonumber\\
- \frac{t'}{2s}\sum\limits_{< ij> }  \bigg[ \left(
{\varphi _{i\sigma}^+ \varphi _{j\sigma} - 2s} \right)\left( {\Psi
_{1i}^{ + A} \Psi _{2j}^A+\Psi _{2i}^{ + A} \Psi _{1j}^A  +
\Psi _{2j}^{ + B} \Psi _{1i}^B+\Psi _{1j}^{ + B} \Psi_{2i}^B}\right)\nonumber\\
+\left( {\varphi _{j\sigma }^+ \varphi_{i\sigma} -2s} \right) \left(
{\Psi _{1i}^{+ B} \Psi _{2j}^B+\Psi _{2i}^{+ B} \Psi _{1j}^B +\Psi
_{2j}^{ + A} \Psi _{1i}^A + \Psi _{1j}^{ + A} \Psi _{2i}^A }
\right)\bigg]\!\Bigg\} \eea \bea \label{S2}\hskip -.3cm S_{\rm
s-f}^{(2)} = \int\limits_{0}^\beta  d\tau \bigg\{ -\frac
{t_1}{2s}\sum\limits_{< ij> } \bigg[\left( {\varphi _{i1}^ + \varphi
_{j2}^ +   - \varphi _{j1}^ +  \varphi _{i2}^ + } \right)\left(
{\Psi _{1j}^{ + A} \Psi _{1i}^B  - \Psi _{1i}^{ + A} \Psi _{1j}^B }
\right) + \left( {\varphi _{i1} \varphi _{j2}  - \varphi _{i2}
\varphi _{j1} }\right) \left( {\Psi
_{1i}^{ + B} \Psi_{1j}^A  - \Psi _{1j}^{ + B} \Psi _{1i}^A } \right)\bigg]\nonumber\\
-\frac {t_2}{2s}\sum\limits_{< ij> }  \bigg[\left( {\varphi _{i1}^ + \varphi _{j2}^ +   - \varphi _{j1}^ +
\varphi
_{i2}^ +  } \right)\left( {\Psi _{2j}^{ + A} \Psi _{2i}^B  - \Psi
_{2i}^{ + A} \Psi _{2j}^B } \right) + \left( {\varphi _{i1}
\varphi _{j2}  - \varphi _{i2} \varphi _{j1} } \right)\left( {\Psi
_{2i}^{ + B} \Psi_{2j}^A  - \Psi _{2j}^{ + B} \Psi _{2i}^A } \right)\bigg] \nonumber\\
-\frac{t'}{2s}\sum\limits_{< ij> }  \bigg[\left( \varphi _{i1}^ + \varphi _{j2}^ +   - \varphi _{j1}^ +
\varphi
_{i2}^+ \right)\left( {\Psi _{1j}^{ + A} \Psi _{2i}^B+\Psi _{2j}^{
+ A} \Psi _{1i}^B -\Psi _{1i}^{ + A} \Psi _{2j}^B - \Psi _{2i}^{ + A} \Psi _{1j}^B } \right)\nonumber\\
+ \left( {\varphi _{i1} \varphi _{j2}  - \varphi _{i2} \varphi _{j1} } \right)\left( {\Psi _{2i}^{ + B} \Psi
_{1j}^A + \Psi_{1i}^{ + B} \Psi _{2j}^A - \Psi _{1j}^{ + B} \Psi
_{2i}^A -\Psi _{2j}^{ + B} \Psi _{1i}^A} \right)\bigg]\!\Bigg\} \eea\end{widetext}

To diagonalize the free fermion Hamiltonian $H_{\rm f}$ we switch to
momentum space, \bea H_{\rm f} & = & \sum\limits_{lk}
\left[\varepsilon _{kl}^A \Psi_{lk}^{A+}\Psi_{lk}^{A}+\varepsilon
_{kl}^B \Psi_{lk}^{B+}\Psi_{lk}^{B}\right]\nonumber \\
& + & \sum\limits_{k \sigma}\left[\varepsilon
_k\left(\Psi_{1k\sigma}^+\Psi_{2k\sigma}+\Psi_{2k\sigma}^+\Psi_{1k\sigma}\right)
\right]\eea where the dispersions for $A$ and $B$ fermions are \be
\begin{array}{l}
\displaystyle \varepsilon _{kl}^A  =  - 2t_l \sum\limits_k {\cos k_\mu  }  - sJ_l - \mu\\[15pt]
\displaystyle \varepsilon _{kl}^B  =  - 2t_l \sum\limits_k {\cos k_\mu  }  + sJ_l - \mu\\[15pt]
\displaystyle\varepsilon(k)=-2t'  \sum\limits_\mu \cos k_\mu
\end{array} \ee The Hamiltonian is diagonalized by means of the transformation: \be
\label{bog-ap}{\hskip -.5cm \begin{array}{l}
   {\Psi _{1k}^A  = u_k^A f_{1k}^A  + v_k^A f_{2k}^A }  \\[15pt]
   {\Psi _{1k}^B  = u_k^B f_{1k}^B  + v_k^B f_{2k}^B }
\end{array}} \hskip .63cm {\begin{array}{l}
   {\Psi _{2k}^A  =  - v_k^A f_{1k}^A  + u_k^A f_{2k}^A }\\[15pt]
   {\Psi _{2k}^B  =  - v_k^B f_{1k}^B  + u_k^B f_{2k}^B }
\end{array}}  \ee

Solving the equations for $u$ and $v$ gives
\bea &&\begin{array}{l} \displaystyle u^R_k  = \sqrt {\frac{1}{2}(1 + x^R_k)}\\[10pt]
\displaystyle v^R_k = {\mathop{ {\rm sign} \left(\varepsilon (k)\right)}}
\sqrt {\frac{1}{2}(1 - x^R_k)}\end{array}\\[5pt]
\textrm{with} \hskip .6cm &&\hskip .1cm x^R_k  = \frac{\varepsilon _{2k}^R - \varepsilon _{1k}^R
}{\sqrt{4\varepsilon ^2 (k) + \left( {\varepsilon _{2k}^R -
\varepsilon _{1k}^R } \right)^2 }} \eea

The resulting Hamiltonian is  \bea H_{\rm f}  =\sum\limits_k \bigg[
E_{1k}^A f_{1k}^{ + A}
f_{1k}^A &+& E_{1k}^B f_{1k}^{ + B} f_{1k}^B\nonumber\\
+\,E_{2k}^A f_{2k}^{ + A} f_{2k}^A  &+& E_{2k}^B f_{2k}^{ + B} f_{2k}^B\, \bigg] \eea
with dispersions for the quasi-particles \be
\begin{array}{l} {\displaystyle E_{1k}^A  = \frac{{\varepsilon _{2k}^A  + \varepsilon _{1k}^A
}}{2} - \frac{1}{2}\sqrt {4\varepsilon ^2 (k) + \left(
{\varepsilon _{2k}^A  - \varepsilon _{1k}^A } \right)^2 }}\\[15pt]
{\displaystyle E_{2k}^A  = \frac{{\varepsilon _{2k}^A  +
\varepsilon _{1k}^A }}{2} + \frac{1}{2}\sqrt {4\varepsilon ^2 (k)
+ \left( {\varepsilon _{2k}^A  - \varepsilon _{1k}^A } \right)^2 }
}\end{array}\ee \be
\begin{array}{l}  {\displaystyle E_{1k}^B  = \frac{{\varepsilon _{2k}^B  +
\varepsilon _{1k}^B }}{2} - \frac{1}{2}\sqrt {4\varepsilon ^2 (k)
+ \left( {\varepsilon _{2k}^B  - \varepsilon _{1k}^B } \right)^2 } } \\[15pt]
{\displaystyle E_{2k}^B  = \frac{{\varepsilon _{2k}^B  +
\varepsilon _{1k}^B }}{2} + \frac{1}{2}\sqrt {4\varepsilon ^2 (k)
+ \left( {\varepsilon _{2k}^B  - \varepsilon _{1k}^B } \right)^2 }
} \end{array} \ee

To write the spin-fermion interaction $S_{\rm s-f}$ in terms of the
new fermions we introduce the notations \be \hskip -.4cm f_{k_1 }^ +
= \left(\! {\begin{array}{*{20}c}
   {f_{1k_1 }^{ + A} } & {f_{1k_1 }^{ + B} } & {f_{2k_1 }^{ + A} } & {f_{2k_1 }^{ + B} }  \\
\end{array}}\!\right)\qquad
f_{k_2 }  = \left(\! {\begin{array}{c}
   {f_{1k_2 }^A }  \\[5pt]
   {f_{1k_2 }^B }  \\[5pt]
   {f_{2k_2 }^A }  \\[5pt]
   {f_{2k_2 }^B }  \\[5pt]
\end{array}}\! \right)\ee
Now we rewrite the action in the form \be S_{\rm s-f} =\!
\int\limits_0^\beta \! {d\tau _1 d\tau _2 \! \sum\limits_{k_1 k_2 }
\! {f_{k_1 }^ + \left( {\tau _1\!} \right)} } W_{k_1 k_2 } \left(
{\tau _1 \! -\! \tau _2 } \right)f_{k_2 }\!\left( {\tau _2\! }
\right) \ee where the matrix $W_{k_1 k_2 } \left( {\tau _1 \! -\!
\tau _2 } \right)$ is a sum of two terms \be\! W_{k_1 k_2 } \left(
\!{\tau _1\!  -\! \tau _2 }\! \right) = W_{k_1 k_2 }^0 \left( {\tau
_1\! - \!\tau _2 } \right) + W_{k_1 k_2 }^{\rm int}\left( {\tau _1\!
-\! \tau _2 } \right). \ee $W_{k_1 k_2 }^0\left(\tau_1 -
\tau_2\right)$ is the free fermion action
\begin{widetext}\be W_{k_1 k_2 }^0 \left( {\tau _1  - \tau _2 }
\right) = \delta _{k_1 k_2 } \delta \left( {\tau _1 - \tau _2 }
\right)\left( {\begin{array}{*{20}c}
   {\frac{\partial }{{\partial \tau _2 }} + E_{1k_2 }^A } & 0 & 0 & 0\\[5pt]
   0 & {\frac{\partial }{{\partial \tau _2 }} + E_{1k_2 }^B } & 0 & 0\\[5pt]
   0 & 0 & {\frac{\partial }{{\partial \tau _2 }} + E_{2k_2 }^A } & 0\\[5pt]
   0 & 0 & 0 & {\frac{\partial }{{\partial \tau _2 }} + E_{2k_2 }^B }
\end{array}} \right)\ee
while $W_{k_1 k_2 }^{\rm int}\left(\tau_1 - \tau_2\right)$ is a sum
of two matrixes , corresponding to $S_{\rm s-f}^{(1)}$ and $S_{\rm
s-f}^{(2)}$ \bea W_{k_1 k_2 }^{{\mathop{\rm int}} } \left( {\tau _1
- \tau _2 } \right) &=&\delta \left( {\tau _1  - \tau _2 }
\right)\left( {\begin{array}{*{20}c}
   {K_{k_1 k_2 } \left( {\tau _2 } \right)} & 0 & 0 & 0 \\[5pt]
   0 & {L_{k_1 k_2 } \left( {\tau _2 } \right)} & 0& 0\\[5pt]
   0 & 0& {N_{k_1 k_2 } \left( {\tau _2 } \right)} &0\\[5pt]
   0 & 0 & 0 & {O_{k_1 k_2 } \left( {\tau _2 } \right)}
\end{array}} \right)\nonumber\\[8pt]
&+& \delta \left( {\tau _1  - \tau _2 } \right)\left(
{\begin{array}{*{20}c}
   0 & {A_{k_1 k_2 } \left( {\tau _2 } \right)} & 0 & {B_{k_1 k_2 } \left( {\tau _2 } \right)}\\[5pt]
   {C_{k_1 k_2 } \left( {\tau _2 } \right)} & 0 & {D_{k_1 k_2 } \left( {\tau _2 } \right)} & 0\\[5pt]
   0 & {E_{k_1 k_2 } \left( {\tau _2 } \right)} & 0 & {F_{k_1 k_2 } \left( {\tau _2 } \right)}\\[5pt]
   {G_{k_1 k_2 } \left( {\tau _2 } \right)} & 0 & {I_{k_1 k_2 } \left( {\tau _2 } \right)} & 0
\end{array}} \right)\label{matrix}
\eea

with matrix elements \bea &\displaystyle K_{k_1 k_2 } \left( {\tau
_2 } \right) = -\frac{1}{{2s}}\frac{1}{N}\sum\limits_k \left(
{\sum\limits_{\mu  = 1}^3 {\frac{{\cos k_\mu  }}{3}} } \right)
\sum\limits_{ < ij > } \left( {\varphi _{i\sigma }^ + \varphi
_{j\sigma } + \varphi _{j\sigma }^ +  \varphi _{i\sigma }
- 4s} \right) \left[ t_1 \left(u_{k }^A\right)^2 +t_2\left(v_k^A\right)^2 -2t'u_k^A v_k^A  \right]\nonumber \\[5pt]
&\displaystyle L_{k_1 k_2 } \left( {\tau _2 } \right) =
-\frac{1}{{2s}}\frac{1}{N}\sum\limits_k \left( {\sum\limits_{\mu  =
1}^3 {\frac{{\cos k_\mu  }}{3}} } \right) \sum\limits_{ < ij > }
\left( {\varphi _{i\sigma }^ +  \varphi _{j\sigma } + \varphi
_{j\sigma }^ +  \varphi _{i\sigma }
- 4s} \right) \left[ t_1 \left(u_{k }^B\right)^2 +t_2\left(v_k^B\right)^2 -2t'u_k^B v_k^B  \right]\nonumber \\[-5pt]
\\[-5pt]
&\displaystyle K_{k_1 k_2 } \left( {\tau _2 } \right) =
 -\frac{1}{{2s}}\frac{1}{N}\sum\limits_k \left( {\sum\limits_{\mu  = 1}^3 {\frac{{\cos k_\mu  }}{3}} } \right)
 \sum\limits_{ < ij > } \left( {\varphi _{i\sigma }^ +  \varphi _{j\sigma } +
 \varphi _{j\sigma }^ +  \varphi _{i\sigma }
- 4s} \right) \left[ t_1 \left(v_{k }^A\right)^2 +t_2\left(u_k^A\right)^2 +2t'u_k^A v_k^A  \right] \nonumber\\[5pt]
&\displaystyle L_{k_1 k_2 } \left( {\tau _2 } \right) =
 -\frac{1}{{2s}}\frac{1}{N}\sum\limits_k \left( {\sum\limits_{\mu  = 1}^3 {\frac{{\cos k_\mu  }}{3}} }
 \right) \sum\limits_{ < ij > } \left( {\varphi _{i\sigma }^ +  \varphi _{j\sigma } +
 \varphi _{j\sigma }^ +  \varphi _{i\sigma }
- 4s} \right) \left[ t_1 \left(v_{k }^B\right)^2
+t_2\left(u_k^B\right)^2 +2t'u_k^B v_k^B  \right]\nonumber \eea

\bea &\displaystyle A_{k_1 k_2 } \left( {\tau _2 } \right) =  -
\frac{1}{{2s}}\frac{1}{N}\sum\limits_{ < ij > } {\left( {\varphi
_{i1}^ +  \varphi _{j2}^ +   - \varphi _{j1}^ +  \varphi _{i2}^ + }
\right)} \left( {e^{ - ik_1 r_j  + ik_2 r_i }  - e^{ - ik_1 r_i +
ik_2 r_j } } \right)\left[ {t_1 u_{k_1 }^A u_{k_2 }^B  + t_2 v_{k_1
}^A v_{k_2 }^B  - t'u_{k_1 }^A v_{k_2 }^B  - t'v_{k_1 }^A
u_{k_2 }^B } \right]\nonumber \\[5pt]
&\displaystyle B_{k_1 k_2 } \left( {\tau _2 } \right) =  -
\frac{1}{{2s}}\frac{1}{N}\sum\limits_{ < ij > } {\left( {\varphi
_{i1}^ +  \varphi _{j2}^ +   - \varphi _{j1}^ +  \varphi _{i2}^ + }
\right)} \left( {e^{ - ik_1 r_j  + ik_2 r_i }  - e^{ - ik_1 r_i +
ik_2 r_j } } \right)\left[ {t_1 u_{k_1 }^A v_{k_2 }^B  - t_2 v_{k_1
}^A u_{k_2 }^B  + t'u_{k_1 }^A u_{k_2 }^B  - t'v_{k_1 }^A
v_{k_2 }^B } \right]\nonumber \\[-5pt]
\\[-3pt]
&\displaystyle E_{k_1 k_2 } \left( {\tau _2 } \right) =  -
\frac{1}{{2s}}\frac{1}{N}\sum\limits_{ < ij > } {\left( {\varphi
_{i1}^ +  \varphi _{j2}^ +   - \varphi _{j1}^ +  \varphi _{i2}^ + }
\right)} \left( {e^{ - ik_1 r_j  + ik_2 r_i }  - e^{ - ik_1 r_i +
ik_2 r_j } } \right)\left[ {t_1 v_{k_1 }^A u_{k_2 }^B  - t_2 u_{k_1
}^A v_{k_2 }^B  - t'v_{k_1 }^A v_{k_2 }^B  + t'u_{k_1 }^A
u_{k_2 }^B } \right]\nonumber \\[5pt]
&\displaystyle F_{k_1 k_2 } \left( {\tau _2 } \right) = -
\frac{1}{{2s}}\frac{1}{N}\sum\limits_{ < ij > } {\left( {\varphi
_{i1}^ +  \varphi _{j2}^ +   - \varphi _{j1}^ +  \varphi _{i2}^ + }
\right)} \left( {e^{ - ik_1 r_j  + ik_2 r_i }  - e^{ - ik_1 r_i +
ik_2 r_j } } \right)\left[ {t_1 v_{k_1 }^A v_{k_2 }^B  + t_2 u_{k_1
}^A u_{k_2 }^B  + t'v_{k_1 }^A u_{k_2 }^B  + t'u_{k_1 }^A v_{k_2 }^B
} \right]\nonumber \eea

\bea &\displaystyle C_{k_1 k_2 } \left( {\tau _2 } \right) =  -
\frac{1}{{2s}}\frac{1}{N}\sum\limits_{ < ij > } {\left( {\varphi
_{i1} \varphi _{j2}  - \varphi _{j1} \varphi _{i2} } \right)} \left(
{e^{ - ik_1 r_i  + ik_2 r_j }  - e^{ - ik_1 r_j  + ik_2 r_i } }
\right)\left[ {t_1 u_{k_1 }^B u_{k_2 }^A  + t_2 v_{k_1 }^B v_{k_2
}^A  - t'u_{k_1 }^B v_{k_2 }^A  - t'v_{k_1 }^B u_{k_2 }^A }
\right] \nonumber\\[5pt]
&\displaystyle D_{k_1 k_2 } \left( {\tau _2 } \right) =  -
\frac{1}{{2s}}\frac{1}{N}\sum\limits_{ < ij > } {\left( {\varphi
_{i1} \varphi _{j2}  - \varphi _{j1} \varphi _{i2} } \right)} \left(
{e^{ - ik_1 r_i  + ik_2 r_j }  - e^{ - ik_1 r_j  + ik_2 r_i } }
\right)\left[ {t_1 u_{k_1 }^B v_{k_2 }^A  - t_2 v_{k_1 }^B u_{k_2
}^A  + t'u_{k_1 }^B u_{k_2 }^A  - t'v_{k_1 }^B v_{k_2 }^A }
\right] \nonumber\\[-5pt]
\\[-3pt]
&\displaystyle G_{k_1 k_2 } \left( {\tau _2 } \right) =  -
\frac{1}{{2s}}\frac{1}{N}\sum\limits_{ < ij > } {\left( {\varphi
_{i1} \varphi _{j2}  - \varphi _{j1} \varphi _{i2} } \right)} \left(
{e^{ - ik_1 r_i  + ik_2 r_j }  - e^{ - ik_1 r_j  + ik_2 r_i } }
\right)\left[ {t_1 v_{k_1 }^B u_{k_2 }^A  - t_2 u_{k_1 }^B v_{k_2
}^A  - t'v_{k_1 }^B v_{k_2 }^A  + t'u_{k_1 }^B u_{k_2 }^A }
\right]\nonumber \\[5pt]
&\displaystyle I_{k_1 k_2 } \left( {\tau _2 } \right) =  -
\frac{1}{{2s}}\frac{1}{N}\sum\limits_{ < ij > } {\left( {\varphi
_{i1} \varphi _{j2}  - \varphi _{j1} \varphi _{i2} } \right)} \left(
{e^{ - ik_1 r_i  + ik_2 r_j }  - e^{ - ik_1 r_j  + ik_2 r_i } }
\right)\left[ {t_1 v_{k_1 }^B v_{k_2 }^A  + t_2 u_{k_1 }^B u_{k_2
}^A  + t'v_{k_1 }^B u_{k_2 }^A  + t'u_{k_1 }^B v_{k_2 }^A }
\right]\nonumber \eea\end{widetext}where $N$ is the number of
lattice's sites. Integrating the fermions out we obtain the
effective action $S_{\rm eff}$
\be S_{\rm eff}  =  - \ln \det W =  - {\mathop{\rm Tr}\nolimits} \ln
W \ee Using the properties of the logarithm \bea {\mathop{\rm
Tr}\nolimits} \ln W = {\mathop{\rm Tr}\nolimits}
\ln \left( {W^0  + W^{{\mathop{\rm int}} } } \right) = {\mathop{\rm Tr}\nolimits} \ln W^0 \nonumber\\[5pt]
+{\mathop{\rm Tr}\nolimits} \ln \left( {\mathbbm{1} + \left( {W^0 }
\right)^{ - 1} W^{{\mathop{\rm int}} } } \right) \eea we rewrite the
effective action in the form \be S_{\rm eff}  =  - {\mathop{\rm
Tr}\nolimits} \ln \left( {1 + \left( {W^0 } \right)^{ - 1}
W^{{\mathop{\rm int}} } } \right)\ee where the term ${\mathop{\rm
Tr}\nolimits} \ln W^0$ doesn't depend on the Schwinger bosons and we
have  dropped it. Finally, we expand the effective action in powers
of \be \hskip -.3cm X_{k_1 k_2 } \left( {\tau _1 ,\tau _2 }
\right)\! =\! \sum\limits_q \!{\int\!\! {d\tau \!\!\left[ {W_{k_1
q}^0 \left( {\tau _1 ,\tau } \right)} \right]^{ - 1}\! W_{qk_2
}^{{\mathop{\rm int}} }\! \left( {\tau ,\tau _2 } \right)}
}\label{X} \ee The result is \be\label{X1}  S_{\rm eff}= -
{\mathop{\rm Tr}\nolimits} X + \frac{1}{2}{\mathop{\rm Tr}\nolimits}
X^2  + \ldots \ee The inverse matrix $\left(W_{k_1
k_2}^0\right)^{-1}$ is given by
\begin{widetext}\be \left( {W_{k_1 k_2 }^0 } \right)^{ -
1} \left( {\tau _1 ,\tau _2 } \right) = \left(
{\begin{array}{*{20}c}
   {\delta _{k_1 k_2 } S_{1k_1 }^A \left( {\tau _1  - \tau _2 } \right)} & 0 & 0 & 0\\[5pt]
   0 & {\delta _{k_1 k_2 } S_{1k_1 }^B \left( {\tau _1  - \tau _2 } \right)} & 0 & 0\\[5pt]
   0 & 0 & {\delta _{k_1 k_2 } S_{2k_1 }^A \left( {\tau _1  - \tau _2 } \right)} & 0\\[5pt]
   0 & 0 & 0 & {\delta _{k_1 k_2 } S_{2k_1 }^B \left( {\tau _1  - \tau _2 } \right)}
\end{array}} \right) \ee\end{widetext}
where $S_{lk}^\sigma  \left( \omega  \right) =\displaystyle
\frac{1}{{ - i\omega + E_{lk}^\sigma  }}$ ($\sigma = A\; \rm or\;
B$, $l=1\; \rm or\; 2$). Replacing \eqref{matrix} into \eqref{X}, we
end up with two terms for $X_{k_1 k_2}$, one which is diagonal
$X_{k_1 k_2}^{(1)}$, and one with zero diagonal elements $X_{k_1
k_2}^{(2)}$. Hence, one obtains for the trace of the matrix $X$
\be\mathop{\rm Tr} X=\mathop{\rm Tr} X^{(1)}.\ee where
\begin{widetext}\be X_{k_1 k_2 }^{(1)} \left( {\tau _1 ,\tau _2 }
\right) \!=\! \left(\! {\begin{array}{*{4}c} {S_{1k_1 }^A \!
\left( {\tau _1 \!-\! \tau _2 } \right)\!K_{k_1 k_2 } \left( {\tau _2 } \right)}& 0 & 0 & 0  \\[5pt]
0 & {S_{1k_1 }^B \!\left( {\tau _1 \!-\! \tau _2 } \right)\!L_{k_1 k_2 } \left( {\tau _2 } \right)} & 0 & 0\\[5pt]
0 & 0 & {S_{2k_1 }^A \!\left( {\tau _1 \!-\! \tau _2 } \right)\!N_{k_1 k_2 } \left( {\tau _2 } \right)} & 0\\[5pt]
0 & 0 & 0& {S_{2k_1 }^B \!\left( {\tau _1 \!-\! \tau _2 }
\right)\!O_{k_1 k_2 } \left( {\tau _2 }\right)}
\end{array}}\!\right) \ee
and the first term in the effective action \eqref{X1} is
\bea\nonumber &&S_{\rm eff}^{(1)} =-
\frac{1}{{2s}}\frac{1}{N}\sum\limits_k \left( {\sum\limits_{\mu  =
1}^d {\frac{{\cos k_\mu  }}{d}} } \right) \int\limits_0^\beta d\tau
\sum\limits_{ < ij > } \left( {\varphi _{i\sigma }^ +  \varphi
_{j\sigma } + \varphi _{j\sigma }^ + \varphi _{i\sigma } - 4s}
\right)\Bigg\{2t'\bigg[ u_k^A v_k^A \left( {n_{2k}^A - n_{1k}^A }
\right)  +u_k^B v_k^B \left( {n_{2k}^B - n_{1k}^B}  \right)\bigg]+\\
&&+ t_1 \bigg[\!\left(u_k^A\right)^2 n_{1k}^A +  \left(v_k^A\right)^2 n_{2k}^A  + \left(u_k^B\right)^2 n_{1k}^B +
\left(v_k^B\right)^2 n_{2k}^B  \bigg]\! +t_2 \bigg[\!\left(u_k^A\right)^2 n_{2k}^A +
\left(v_k^A\right)^2 n_{1k}^A  + \left(u_k^B\right)^2 n_{2k}^B +
\left(v_k^B\right)^2 n_{1k}^B\bigg]\!\Bigg\}\eea\end{widetext}

To calculate the contribution of $S_{\rm eff}^{(1)}$ to the
spin-stiffness constant $\rho$ in \eqref{eff} we use the
Holstein-Primakoff representation for the Schwinger bosons
\bea\label{holstein} & & \varphi _{1i} = \varphi _{1i}^ + = \sqrt
{2s - \frac sM a_i^ +  a_i }\\ & & \varphi _{2i} = \sqrt\frac sM\,
a_i \qquad \varphi _{2i}^ +   = \sqrt\frac sM \, a_i^ +  \nonumber
\eea and keep the terms quadratic with respect to the magnons
$a_i,a^+_i$. The result is
\begin{widetext}
\bea\label{rho1}\nonumber \rho_1=
\frac{1}{{2M}}\frac{1}{N}\sum\limits_k \left( {\sum\limits_{\mu  =
1}^d {\frac{{\cos k_\mu  }}{d}} } \right)\Bigg\{2t'\bigg[  \Big(
u_k^A v_k^A \left( {n_{2k}^A - n_{1k}^A }
\right)+u_k^B v_k^B \left( {n_{2k}^B - n_{1k}^B}  \right)\Big) \bigg]\\
+t_1 \bigg[\left(u_k^A\right)^2 n_{1k}^A\! + \!\left(v_k^A\right)^2
n_{2k}^A  \!+\! \left(u_k^B\right)^2 n_{1k}^B \!+\!
\left(v_k^B\right)^2 n_{2k}^B  \bigg]\! +\!t_2
\bigg[\left(u_k^A\right)^2 n_{2k}^A \!+\! \left(v_k^A\right)^2
n_{1k}^A\! +\! \left(u_k^B\right)^2 n_{2k}^B \!+\!
\left(v_k^B\right)^2 n_{1k}^B\bigg]\!\Bigg\}\eea
\end{widetext}

Calculating the contribution of the second term in \eqref{X1}  to
the effective hamiltonian \eqref{eff} we account for the fact that
$X^{1}$ matrix is quadratic with respect to magnons, hence it
doesn't contribute. The contribution comes from $S_{\rm s-f}^{(2)} $
\eqref{S2} which leads to the matrix $X^{(2)}$.
\begin{widetext}\be X_{k_1 k_2 }^{(2)} \left( {\tau _1 ,\tau _2 }
\right) = \left( {\begin{array}{*{20}c} 0 & {S_{1k_1 }^A \left(
{\tau _1 \!-\! \tau _2 } \right)A_{k_1 k_2 } \left( {\tau _2 }
\right)} & 0 & {S_{1k_1 }^A \left( {\tau _1  \!-\! \tau _2 }
\right)B_{k_1 k_2 }\left( {\tau _2 } \right)}  \\[5pt]
{S_{1k_1 }^B \left( {\tau _1  \!-\! \tau _2 } \right)C_{k_1 k_2 }
\left( {\tau _2 } \right)} & 0 & {S_{1k_1 }^B \left( {\tau _1  \!-\!
\tau _2 } \right)D_{k_1 k_2 } \left( {\tau _2 } \right)} & 0\\[5pt]
0 & {S_{2k_1 }^A \left( {\tau _1  \!-\! \tau _2 } \right)E_{k_1 k_2 }
\left( {\tau _2 } \right)} & 0 & {S_{2k_1 }^A \left( {\tau _1  \!-\!
\tau _2 } \right)F_{k_1 k_2 } \left( {\tau _2 } \right)}  \\[5pt]
{S_{2k_1 }^B \left( {\tau _1  \!-\! \tau _2 } \right)G_{k_1 k_2 }
\left( {\tau _2 } \right)} & 0 & {S_{2k_1 }^B \left( {\tau _1  \!-\!
\tau _2 } \right)I_{k_1 k_2 } \left( {\tau _2 } \right)} & 0
\end{array}} \right) \ee\end{widetext}
After some algebra we arrive at the following representation of the
second term in \eqref{X1} \begin{widetext}\bea\label{Seff2}
&&\nonumber \hskip -.2cm S_{\rm eff}^{(2)} =\!\! \int\! d\tau _1
d\tau _2 \!\! \sum\limits_{k_1 k_2 } \Big[ S_{1k_1 }^A \left( {\tau
_1 - \tau _2 } \right)A_{k_1 k_2 } \left( {\tau _2 } \right)S_{1k_2
}^B \left( {\tau _2  - \tau _1 } \right)C_{k_2 k_1 } \left( {\tau _1
} \right) + S_{1k_1 }^A \left( {\tau _1  - \tau _2 } \right)B_{k_1
k_2 } \left( {\tau _2 } \right)S_{2k_2 }^B \left( {\tau _2  - \tau
_1 }
\right)G_{k_2 k_1 } \left( {\tau _1 } \right)\\[5pt]
&&+ S_{2k_1 }^A \left( {\tau _1  - \tau _2 } \right)E_{k_1 k_2 }
\left( {\tau _2 } \right)S_{1k_2 }^B \left( {\tau _2  - \tau _1 }
\right)D_{k_2 k_1} \left( {\tau _1 } \right)+S_{2k_1 }^A \left( {\tau _1  - \tau _2 } \right)
F_{k_1 k_2 } \left( {\tau _2 }
\right)S_{2k_2 }^B \left( {\tau _2  - \tau _1 } \right)I_{k_2 k_1
} \left( {\tau _1 } \right) \Big] \eea\end{widetext}

Switching from imaginary time $\tau$ representation to frequency
$\omega$ representation we calculate the expressions in small
$\omega$ approximation. The result is a \be S_{lk_1 }^A \left( {\tau
_1 \!-\! \tau _2 } \right)S_{l'k_2 }^B \left( {\tau _1 \!-\!\tau _2
} \right) \simeq \delta \left( {\tau _1 \!-\! \tau _2 }
\right)\!\int \!{\frac{{d\omega }}{{2\pi }}} S_{lk_1 }^A (\omega
)S_{l'k_2 }^B (\omega ).\ee Next we make a change of wave-vectors
variables
$k_1 = q +\frac 12 k$, $k_1= q -\frac 12 k$, and calculate the
expressions in small wave-vector $k$ approximation. The expression
\eqref{Seff2} calculated in small frequency and small wave-vector
approximation has the form
\begin{widetext}\bea & & S_{\rm eff}^{(2)}=\int\frac{d\omega}{2\pi s^2 }
\frac{1}{N}\sum\limits_q \left(\sum\limits_{\mu=1}^3 \frac{\sin^2
q_\mu}{3}\right) \sum\limits_{ij }\Bigg[{\left( {\varphi _{i1}^ +
\varphi _{j2}^ + - \varphi _{j1}^ + \varphi _{i2}^ + }
\right)} \left( {\varphi _{i1} \varphi_{j2}  - \varphi _{j1} \varphi _{i2} } \right)\Bigg] \\
&&\Big[ S_{1q}^A (\omega )S_{1q}^B (\omega )\left( {t_1 u_q^A u_q^B
+ t_2 v_q^A v_q^B  - t'u_q^A v_q^B  - t'v_q^A u_q^B } \right)^2
+S_{1q}^A (\omega )S_{2q}^B
(\omega )\left( {t_1 u_q^A v_q^B  - t_2 v_q^A u_q^B  + t'u_q^A u_q^B  - t'v_q^A v_q^B } \right)^2  \nonumber\\
& + & S_{2q}^A (\omega )S_{1q}^B (\omega )\left( {t_1 v_q^A u_q^B  -
t_2 u_q^A v_q^B  - t'v_q^A v_q^B  + t'u_q^A u_q^B } \right)^2
+S_{2q}^A (\omega )S_{2q}^B (\omega )\left( {t_1 v_q^A v_q^B  + t_2
u_q^A u_q^B  + t'v_q^A u_q^B  + t'u_q^A v_q^B } \right)^2
\Big]\nonumber \eea\end{widetext} Our third step is to express the
products of the Green functions, in the above equation, in terms of
the fermi function $n(E)=1/(e^{E}+1)$ \be \int {\frac{{d\omega
}}{{2\pi }}S_{1q}^A (\omega )S_{1q}^B (\omega )}  = \frac{{n\left(
{E_{1q}^B } \right) - n\left( {E_{1q}^A } \right)}}{\lower 3pt
\hbox{$E_{1q}^B - E_{1q}^A $}}\ee \be \int {\frac{{d\omega }}{{2\pi
}}S_{1q}^A (\omega )S_{2q}^B (\omega )} = \frac{{n\left( {E_{2q}^B }
\right) - n\left( {E_{1q}^A } \right)}}{\lower 3pt\hbox{$E_{2q}^B  -
E_{1q}^A$}}\ee \be \int {\frac{{d\omega }}{{2\pi }}S_{2q}^A (\omega
)S_{1q}^B (\omega )}  = \frac{{n\left( {E_{1q}^B } \right) - n\left(
{E_{2q}^A } \right)}}{\lower 3 pt \hbox{$E_{1q}^B  - E_{2q}^A $}}\ee
\be \int {\frac{{d\omega }}{{2\pi}}S_{2q}^A (\omega )S_{2q}^B
(\omega )}  = \frac{{n\left( {E_{2q}^B } \right) - n\left( {E_{2q}^A
} \right)}}{\lower 3 pt \hbox{$E_{2q}^B  - E_{2q}^A$}}.\ee Finally
we use the representation of the Schwinger bosons \eqref{holstein}.
To calculate the contribution to the spin-stiffness constant it is
enough to keep only the quadratic terms with respect to magnons
\begin{widetext}\bea\label{rho2} \nonumber&\displaystyle \rho_2 = \frac{2}{M} \frac{1}{V} \sum\limits_q
\left(\sum\limits_{\mu=1}^d \frac{\sin^2
q_\mu}{d}\right)\Bigg[\left( {t_1 u_q^A u_q^B + t_2 v_q^A v_q^B  -
t'u_q^A v_q^B  - t'v_q^A u_q^B } \right)^2  \left(\frac{{n\left(
{E_{1q}^B } \right) - n\left( {E_{1q}^A } \right)}}{\lower 3 pt \hbox{$E_{1q}^B  - E_{1q}^A$}}\right)  +\\
&\nonumber\displaystyle + \left( {t_1 u_q^A v_q^B  - t_2 v_q^A u_q^B  +
t'u_q^A u_q^B  - t'v_q^A v_q^B } \right)^2  \left(\frac{{n\left(
{E_{2q}^B } \right) - n\left( {E_{1q}^A } \right)}}{\lower 3 pt \hbox{$E_{2q}^B  - E_{1q}^A$}}\right) +  \\
&\nonumber\displaystyle + \left( {t_1 v_q^A u_q^B  - t_2 u_q^A v_q^B -
t'v_q^A v_q^B  + t'u_q^A u_q^B } \right)^2 \left(\frac{{n\left(
{E_{1q}^B } \right) - n\left( {E_{2q}^A } \right)}}{\lower 3 pt \hbox{$E_{1q}^B  - E_{2q}^A$ }}\right) +\\
&\displaystyle + \left( {t_1 v_q^A v_q^B  + t_2 u_q^A u_q^B  +
t'v_q^A u_q^B  + t'u_q^A v_q^B } \right)^2 \left(\frac{{n\left(
{E_{2q}^B } \right) - n\left( {E_{2q}^A } \right)}}{\lower 3 pt \hbox{$E_{2q}^B  -
E_{2q}^A$ }}\right) \Bigg] \eea\end{widetext}

The spin-stiffness constant in the effective action \eqref{eff} is a
sum of the expressions \eqref{rho1} and \eqref{rho2}
\begin{widetext}\bea\label{rho} \rho = \nonumber \Bigg\{\frac{{t_1
}} {{2M}} \frac{1}{V} \sum\limits_k {\left( {\sum\limits_{\mu  =
1}^d {\frac{{\cos k_\mu }}{d}} } \right)} \left[ \left( {u_k^A }
\right)^2 n_{1k}^A  + \left( {v_k^A } \right)^2 n_{2k}^A + \left(
{u_k^B } \right)^2 n_{1k}^B + \left( {v_k^B } \right)^2 n_{2k}^B\right] + \\
\nonumber + \frac{{t_{2} }}{{2M}}\frac{1}{V}\sum\limits_k
{\left( {\sum\limits_{\mu  = 1}^d {\frac{{\cos k_\mu  }}{d}} }
\right)} \left[ \left( {u_k^A } \right)^2 n_{2k}^A + \left( {v_k^A
} \right)^2 n_{1k}^A + \left( {u_k^B } \right)^2 n_{2k}^B +
\left( {v_k^B } \right)^2 n_{1k}^B \right] +  \\
\nonumber \hskip -1cm + \frac{{t' }} {{M}} \frac{1}{V}
\sum\limits_k {\left( {\sum\limits_{\mu  = 1}^d {\frac{{\cos k_\mu
}}{d}} } \right)} \Big[ {u_k^A v_k^A \left( {n_{2k}^A - n_{1k}^A}
\right) + u_k^B v_k^B \left( n_{2k}^B - n_{1k}^B \right)} \Big] + \\
\nonumber + \frac{2}{M} \frac{1}{V} \sum\limits_k
\left(\sum\limits_{\mu=1}^d \frac{\sin^2 k_\mu} {d} \right) \Bigg[
\left( {t_1 u_k^A u_k^B + t_2 v_k^A v_k^B  - t'u_k^A v_k^B  -
t'v_k^A u_k^B } \right)^2  \left(\frac{n_{1k}^B - n_{1k}^A}{{E_{1k}^B  - E_{1k}^A }}\right)  +\\
\nonumber + \left( {t_1 u_k^A v_k^B  - t_2 v_k^A u_k^B  +
t'u_k^A u_k^B  - t'v_k^A v_k^B } \right)^2  \left(\frac{{n_{2k}^B
- n_{1k}^A }}{{E_{2k}^B  - E_{1k}^A }}\right) +  \\
\nonumber + \left( {t_1 v_k^A u_k^B  - t_2 u_k^A v_k^B -
t'v_k^A v_k^B  + t'u_k^A u_k^B } \right)^2
\left(\frac{{n_{1k}^B - n_{2k}^A}}{{E_{1k}^B  - E_{2k}^A }}\right) +\\
 + \left( {t_1 v_k^A v_k^B  + t_2 u_k^A u_k^B  +
t'v_k^A u_k^B  + t'u_k^A v_k^B } \right)^2 \left(\frac{{n_{2k}^B -
n_{2k}^A}}{{E_{2k}^B  - E_{2k}^A }}\right) \Bigg]\Bigg\} \eea\end{widetext}

\end{document}